\documentclass[pra,twocolumn,showpacs,superscriptaddress,floatfix, nofootinbib]{revtex4-2}
\usepackage[caption=false]{subfig}
\usepackage{amsmath,graphicx}
\usepackage{amsmath,amssymb,mathrsfs,esint}
\usepackage{multirow}
\usepackage{algorithm}
\usepackage[noend]{algpseudocode}
\usepackage{comment}
\usepackage{tabularx}
\usepackage{bm}
\usepackage[normalem]{ulem}
\usepackage[bookmarks=true,
   colorlinks=true,
   linkcolor=blue,
   urlcolor=blue,
   citecolor=blue,
   bookmarks=true,
   hyperindex=true
]{hyperref}

\usepackage{xcolor}
\colorlet{RED}{red}

\begin{document}

\title{Realizing quantum gates  with optically-addressable $^{171}$Yb$^{+}$ ion qudits}

\author{M.A. Aksenov}
\affiliation{P.N. Lebedev Physical Institute of the Russian Academy of Sciences, Moscow 119991, Russia}
\affiliation{Russian Quantum Center, Skolkovo, Moscow 121205, Russia}

\author{I.V. Zalivako}
\affiliation{P.N. Lebedev Physical Institute of the Russian Academy of Sciences, Moscow 119991, Russia}
\affiliation{Russian Quantum Center, Skolkovo, Moscow 121205, Russia}

\author{I.A. Semerikov}
\affiliation{P.N. Lebedev Physical Institute of the Russian Academy of Sciences, Moscow 119991, Russia}
\affiliation{Russian Quantum Center, Skolkovo, Moscow 121205, Russia}

\author{A.S. Borisenko}
\affiliation{P.N. Lebedev Physical Institute of the Russian Academy of Sciences, Moscow 119991, Russia}
\affiliation{Russian Quantum Center, Skolkovo, Moscow 121205, Russia}

\author{N.V. Semenin}
\affiliation{P.N. Lebedev Physical Institute of the Russian Academy of Sciences, Moscow 119991, Russia}
\affiliation{Russian Quantum Center, Skolkovo, Moscow 121205, Russia}

\author{P.L. Sidorov}
\affiliation{P.N. Lebedev Physical Institute of the Russian Academy of Sciences, Moscow 119991, Russia}
\affiliation{Russian Quantum Center, Skolkovo, Moscow 121205, Russia}

\author{A.K.~Fedorov}
\affiliation{Russian Quantum Center, Skolkovo, Moscow 121205, Russia}

\author{K.Yu.~Khabarova}
\affiliation{P.N. Lebedev Physical Institute of the Russian Academy of Sciences, Moscow 119991, Russia}
\affiliation{Russian Quantum Center, Skolkovo, Moscow 121205, Russia}

\author{N.N.~Kolachevsky}
\affiliation{P.N. Lebedev Physical Institute of the Russian Academy of Sciences, Moscow 119991, Russia}
\affiliation{Russian Quantum Center, Skolkovo, Moscow 121205, Russia}

\begin{abstract}
The use of multilevel information carriers, also known as qudits, is a promising path for exploring scalability of quantum computing devices.
Here we present a proof-of-principle realization of a quantum processor register that uses optically-addressed $^{171}$Yb$^{+}$ ion qudits in a linear trap.
The rich level structure of $^{171}$Yb$^{+}$ ions allows using the Zeeman sublevels of the quadrupole clock transition at 435.5 nm for efficient and robust qudit encoding.
We demonstrate the realization of the universal set of gates consisting of single-qudit rotations and a two-qudit Mølmer–Sørensen operation with a two-ququart system,
which is formally equivalent to a universal gate-based four-qubit processor.
Our results paves a way towards further studies of more efficient implementations of quantum algorithms with trapped-ion-based processors
and, specifically, exploring properties of $^{171}$Yb$^{+}$ ion qudits.
\end{abstract}

\maketitle

\section{Introduction}

Quantum algorithms offer significant advantages over best-known classical methods in solving certain computational problems ranging from prime factorization to simulating complex systems~\cite{Shor1994,Lloyd1996,Lloyd2009,Lloyd2014-2,Fedorov2022}.
However, in order to run corresponding quantum algorithms in industry-relevant settings one needs to operate with thousands of logical qubits~\cite{Gidney2021,Troyer2017}, which cannot be provided by currently available quantum devices. 
The need of logical qubits requires sizeable increase in fidelities of quantum gate, in particular, two-qubit operations. 
We note that the possibility to achieve computational advantage with noisy processors for practical problems remains questionable~\cite{Babbush2021-4,Bharti2021}.
Recent progress in experiments with controllable quantum many-body systems based on 
superconducting circuits~\cite{Martinis2019,Pan2021-4}, semiconductor quantum dots~\cite{Vandersypen2022,Morello2022,Tarucha2022}, photonic systems~\cite{Pan2020,Lavoie2022}, 
neutral atoms~\cite{Lukin2021,Browaeys2021,Browaeys2020-2,Saffman2022}, 
and trapped ions~\cite{Monroe2017,Blatt2012,Blatt2018} has made it possible to use such systems for tests on quantum advantage~\cite{Martinis2019,Pan2020,Pan2021-4}, 
quantum simulation~\cite{Lukin2021,Browaeys2021,Browaeys2020-2,Monroe2017,Blatt2012}, 
and prototyping quantum algorithms~\cite{Saffman2022,Blatt2018}. 
Although the reported numbers of qubits in quantum processors tend to hundreds 
(for example, 433-qubit superconducting quantum processor has been announced~\cite{IBMroadmap} and 256-atom quantum simulator has been used for optimization~\cite{Lukin2022}), 
the reported gate fidelities of these large-scale processors are lower than in its small-scale counterparts~\cite{Lukin2018-3,Martinis2019}. 
Moreover, the connectivity between individual qubits is typically limited to their nearest neighbours, although there are various proposals to overcome this limitation~\cite{Lukin2021-9}.
These issues reduce the quantum volume (QV)~\cite{Gambetta20192} achieved by these devices, which is a metric for gate-based quantum processors combining the qubit number and gate fidelities.

Being one of the first platforms proposed for quantum computing~\cite{CiracZoller1995,Wineland1995}, 
today trapped-ion systems show the highest QV of 32768 in experiments by Quantinuum with the 15-qubit H1-1 processor~\cite{Quantinuum2022}.
The trapped ion quantum processor also admits an effective error correction~\cite{Wineland2004,Blatt2011,Blatt2020,Monroe2021,Blatt2021-2,Postler2022}, 
e.g., a fault-tolerant entanglement between two logical qubits has been realized~\cite{Postler2022,Ryan-Anderson2022}. 
Main features of trapped ions are long coherence times~\cite{Kim2021}, high-fidelity gates~\cite{Wineland2016}, and essential all-to-all connectivity.
Still, the scalability to large enough numbers of qubits without the decrease of the gate fidelities remains challenging~\cite{Sage2019,Fedorov2022}. 

An interesting feature of certain physical quantum platforms is their multilevel structure, which can be efficiently used for scaling quantum processors with the use of qudits 
--- $d$-level quantum systems~\cite{Farhi1998,Kessel1999,Kessel2000,Kessel2002,Muthukrishnan2000,Nielsen2002,Berry2002,Klimov2003,Bagan2003,Vlasov2003,Clark2004,Leary2006,Ralph2007,White2008,Ionicioiu2009,Ivanov2012,Li2013,Kiktenko2015,Kiktenko2015-2, Song2016,Frydryszak2017,Bocharov2017,Gokhale2019,Pan2019,Low2020,Jin2021,Martinis2009,White2009,Wallraff2012,Mischuck2012,Gustavsson2015,Martinis2014,Ustinov2015, Morandotti2017,Balestro2017,Low2020,Sawant2020,Senko2020,Pavlidis2021,Rambow2021,OBrien2022,Nikolaeva2022}. 
There are two basic approaches: {(i)} encoding several qubits in a single qudit and (ii) the use of additional qudit levels to substitute ancilla qubits in multiqubit gate decompositions~\cite{Barenco1995}
(for example, for the Toffoli gate~\cite{Ralph2007,White2009,Ionicioiu2009,Wallraff2012,Kwek2020,Baker2020,Kiktenko2020,Kwek2021,Galda2021,Gu2021}). 
These two methods can be efficiently combined~\cite{Nikolaeva2021}. 
Both approaches allow one to increase of the QV since a $d$-state qudit can be thought of as $\log_{2}d$ qubits. 
Additionally, certain two-qubit operations can be replaced by single-qudit ones.
While the presence of additional energy levels is used for various quantum computing platforms~\cite{White2009,Wallraff2012,Galda2021,Gustavsson2015,Martinis2014,Ustinov2015,Hill2021,OBrien2022}, 
the use of qudit encoding is especially useful for trapped ions where switching from qubits to qudits is rather simple from experimental point of view.~\cite{Senko2020,Ringbauer2021}.  
Notably, the first realization of two-qubit gates has used two qubits stored in the degrees of freedom of a single trapped ion, i.e. in the qudit setup~\cite{Wineland1995}.

Recently, mutliqudit quantum processors~\cite{Hill2021,Ringbauer2021,OBrien2022}, including the one based on $^{40}$Ca$^{+}$ qudits~\cite{Ringbauer2021}, have been demonstrated. 
In trapped-ion systems, quantum information is typically encoded in metastable states of ions, which are coupled by optical or microwave (mw) fields to perform quantum operations.
In the case of using qudits, for example, the Zeeman structure of the 729 nm optical transition in $^{40}$Ca$^{+}$ ion has been used for the qudit encoding with $d$ up to 7~\cite{Ringbauer2021}.
One may argue that transition frequencies between certain of the sublevels are more susceptible to magnetic field fluctuations in contrast to the one usually chosen as a qubit (see Fig.~\ref{fig:levels}).
However, with the proper magnetic shielding one can significantly reduce the impact of the field noise on the coherence time. 
It is possible to realize a universal gate set consisting of single-qudit and two-qudit operations.
However, finding best-possible conditions for developing scalable qudit-based quantum processors requires additional research. 
Specifically, another promising option for trapped-ion setups is to use $^{171}$Yb$^{+}$, which offers certain advantages~\cite{Sage2019}. 
In particular, the ytterbium ion level structure provides multiple options for encoding quantum information with a high degree of robustness with respect to decoherence~\cite{Quantinuum2022,Ryan-Anderson2022,Kim2021}.

In this work, we present a two-ququart quantum processor using $^{171}$Yb$^{+}$ ions with the quadrupole clock transition at 435.5 nm used for efficient qudit encoding.
This transition is widely used in metrology~\cite{Tamm2000,Kersale2016,Leute2016,Kolachevsky2018}, 
and it has been recently proposed for robust qubit encoding with ytterbium ions~\cite{ZalivakoSemerikov2021}. 
Here we extend this idea for encoding qudits.
We demonstrate the realization of the full set of quantum gates required for implementing algorithms, which consists of single-qudit gates with fidelities ranges from $83\%$ to $89\%$ and the two-qubit operation with $65\pm4\%$ fidelity.
We note that a single two-qudit Mølmer–Sørensen (MS) gate acting on $|0\rangle$ and $|1\rangle$ states in both ions is sufficient to complete a full two-ququart gate set. Although it is possible to add more two-qudit operations acting on other qudit states to the gate set, it is unnecessary from point of view of universality while would significantly complicate a calibration process.
To the best of our knowledge, our work present the first results on manipulating multiqudit system based on  $^{171}$Yb$^{+}$ ions with optical qudits.

\section{Optical $^{171}$Yb$^{+}$ qudits}\label{sec:optqudit}

$^{171}$Yb$^{+}$ ions are one of the main work horses in quantum metrology and quantum computing~\cite{Sage2019}.
These ions are directly laser cooled with 369.5 nm emission on a strong quasi-cyclic $^2S_{1/2}$ $\to$ $^2P_{1/2}$ transition with repumping using $^2D_{3/2}$ $\to$ $^3[3/2]_{1/2}$ transition at 935.2 nm~\cite{Kolachevsky2019,Ejtemaee2019}.
This process can be implemented with widely available semiconductor lasers and does not require frequency conversion.
Since we are interested in $^{171}$Yb$^{+}$ isotope (the nuclear spin equals $I=1/2$), one can realize isotope-selective trap loading with another readily available diode laser at 398.9 nm.

In particular, the transition between hyperfine components of the ground state $^2S_{1/2}(F=0,m_F=0)$ $\to$ $^2S_{1/2}(F=1,m_F=0)$ is widely used as a mw qubit.
Its transition frequency is insensitive (to the first order) to the magnetic field fluctuations. 
The qubit can be readily initialized by optical pumping to the single Zeeman sublevel of the $^2S_{1/2}(F=0)$ manifold.
We note that entanglement of two logical qubits~\cite{Ryan-Anderson2022}, the largest QV~\cite{Quantinuum2022}, and the record coherence times~\cite{Kim2021} have been demonstrated exactly with this type of mw qubit.

In our previous work~\cite{ZalivakoSemerikov2021}, we have proposed the electric quadrupole transition $^2S_{1/2}(F=0,m_F=0)$ $\to$ $^2D_{3/2}(F=2,m_F=0)$ in $^{171}$Yb$^{+}$ to encode an optical qubit.
This transition has a wavelength of 435.5 nm and the natural linewidth of 3 Hz (the corresponding upper level life time equals $\tau$= 53 ms). 
Such encoding combines already mentioned features of ytterbium ions with advantages of optical qubits, mainly, in addressing. 
First, for addressing optical qubits one can use lasers with visible-light wavelength range, whereas for addressing mw qubits ultraviolet lasers are typically required.
This allows using more efficient optical components, for example, ${\rm TeO_2}$ acousto-optical deflectors (AODs), which exhibit much larger scan range. 
Second, in order to perform a two-qubit operation on mw qubits two non co-propagating laser beams are required, while for optical qubits a single beam is enough.
Moreover, in comparison with $^{40}$Ca$^{+}$ optical qubits, 
the proposed transition in $^{171}$Yb$^{+}$ ions exhibits two orders of magnitude smaller sensitivity to the magnetic field (52 Hz/$\mu$T at 500 $\mu$T field offset required for cooling in comparison to 5600 Hz/$\mu$T in $^{40}$Ca$^{+}$ ions),
which leads to lower requirements for magnetic shielding. 
However, exploring these potential advantages of $^{171}$Yb$^{+}$ ions requires additional studies, which are beyond the scope of the present work.

\begin{figure}
\center{\includegraphics[width=0.9\linewidth]{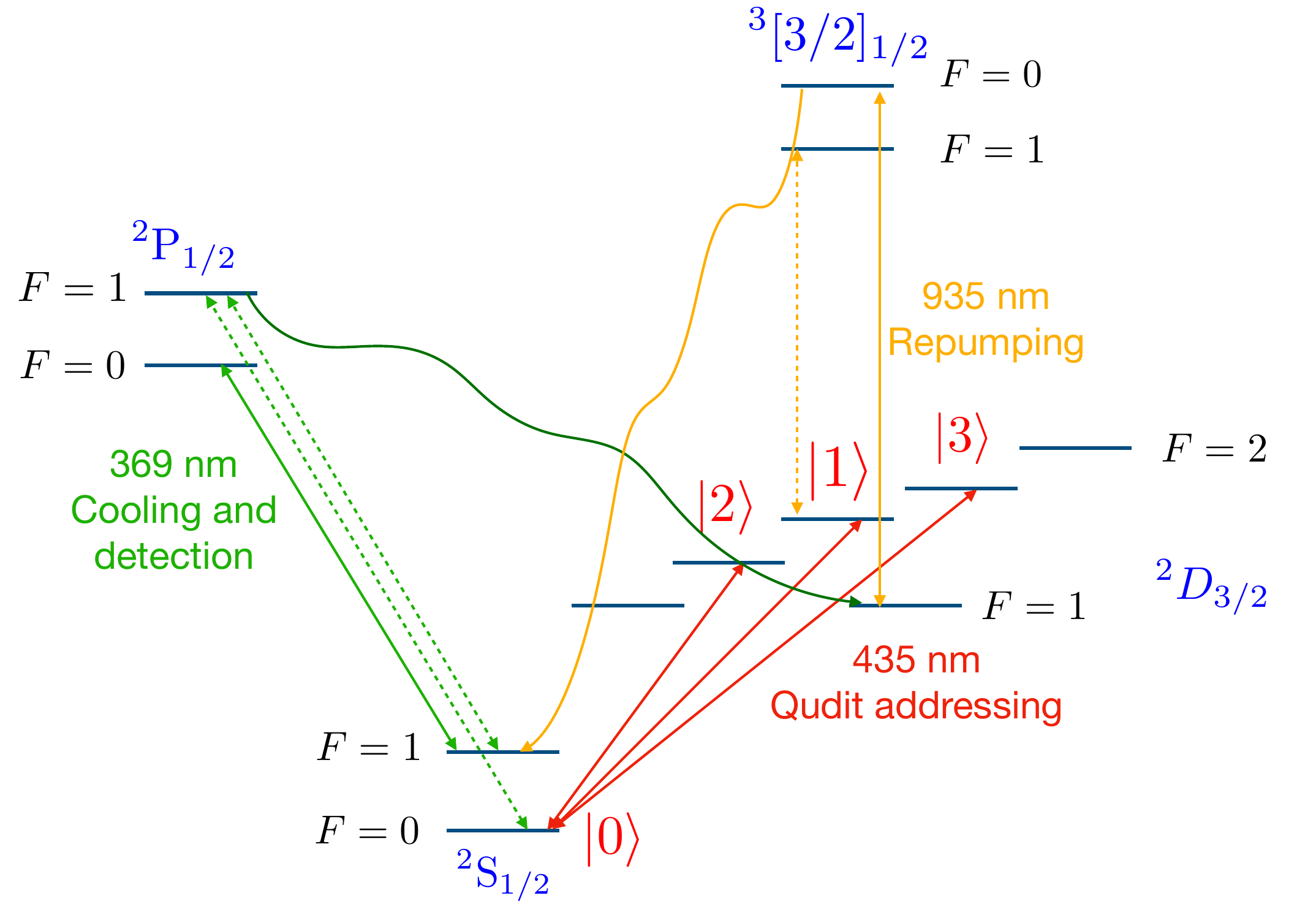}}
	\vskip-3mm
\caption{$^{171}$Yb$^{+}$ level structure used for qudit manipulations: 
the state $|0\rangle$ is coupled to $|1\rangle$, $|2\rangle$, and $|3\rangle$ using single-qudit operations $R_{01}(\phi, \theta)$, $R_{02}(\phi, \theta)$, and $R_{03}(\phi, \theta)$, correspondingly.
Arrows show laser fields used for the ion cooling and readout (green), repumping (orange) and quantum gates (red). 
Dotted arrows show fields generated by phase modulation of the laser beams, while curved lines correspond to spontaneous decays.}
\label{fig:levels}
\end{figure}
 
Similarly to the calcium ions, one may realize optical qudits in $^{171}$Yb$^{+}$. 
All six Zeeman sublevels of both upper and lower levels can be used for the information encoding giving rise to qudits with $d=6$ (see Fig.~\ref{fig:levels}).
However, $^{171}$Yb$^{+}$ qudits are potentially better protected from decoherence due to magnetic field fluctuations, since qudit levels have the integer full angular momentum number $F$, with $F = 0$ for the lower level. 
The $|0\rangle$ and $|1\rangle$ qubit states' energies are sensitive to the magnetic field only in the second order due to zero magnetic quantum numbers. 
Other qudit states can be decoupled from magnetic fields noise using decoupling techniques similar to what is presented in Ref.~\cite{Valahu2022}, 
however, the demonstration of decoupling methods is a topic for further research beyond the scope of the present work.
In our proof-of-principle demonstration we use only four of six available states for ququart encoding ($d = 4$), as this system can be straightforwardly mapped to a conventional qubit-based processor by considering each ququart to encode two qubits.

\begin{figure}
\center{\includegraphics[width=1\linewidth]{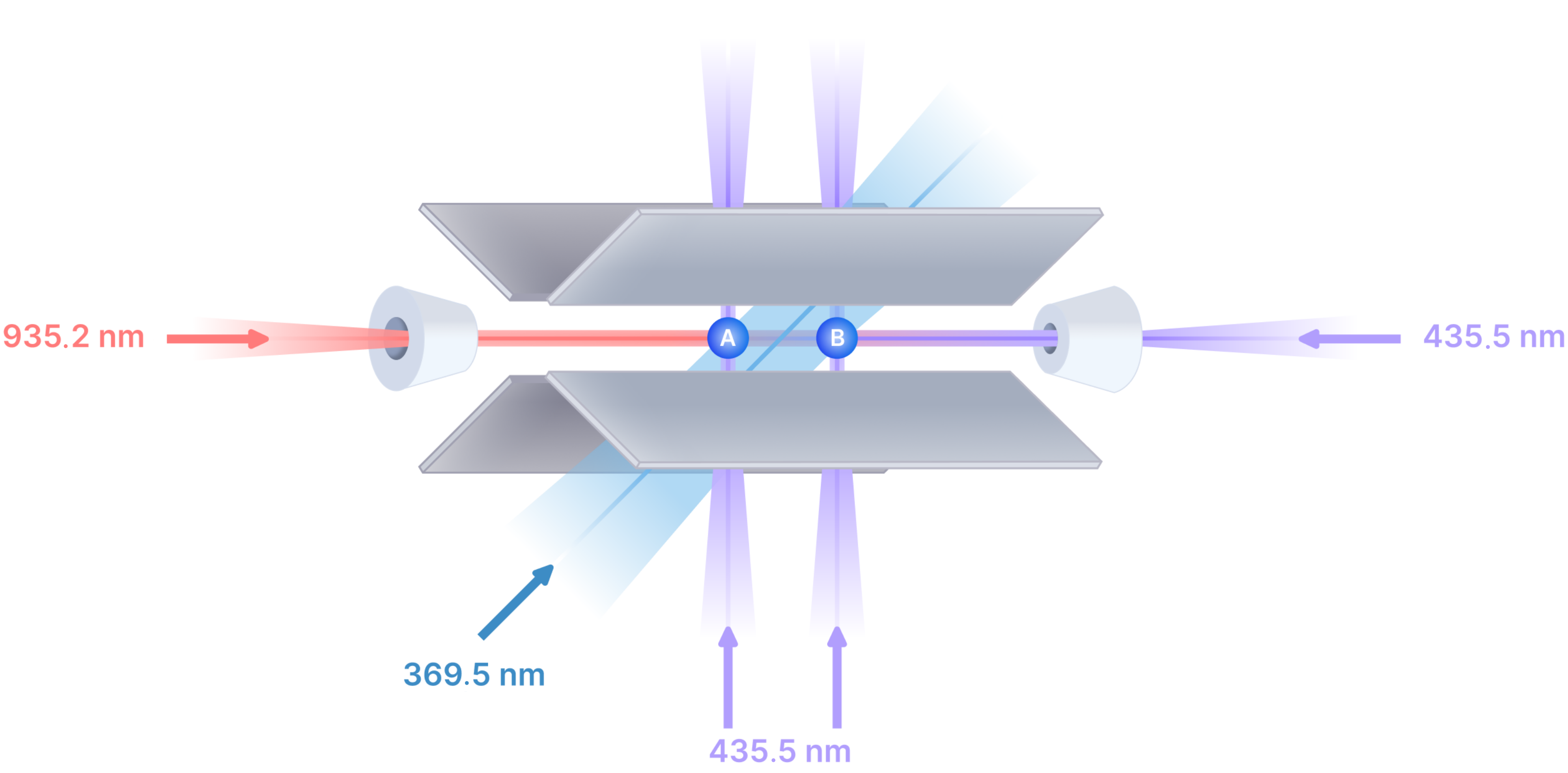}}
	\vskip-3mm
\caption{The one-segment linear 4-blades Paul trap is used for trapping $^{171}$Yb$^{+}$ ions (two ions are labeled as $A$ and $B$).
Laser beams shown in blue (369.5 nm) and red (935.2 nm) are used for Doppler cooling and repumping, respectively. 
The beam at 435.5 nm along the trap axis provides ground state cooling and two-qubit gates. 
Another beam at 435.5 nm is orthogonal to the trap axis and its position can be tuned to address individual ions and perform single-qudit gates.}
\label{fig:trap}
\end{figure}

\subsection{Qudit gates}

In the further discussion the ququart states are denoted as $|0\rangle$, $|1\rangle$, $|2\rangle$, and $|3\rangle$ according to Fig.~\ref{fig:levels}. 
We realize a universal set of gates consisting of single-qudit operation and a single two-qudit gate.
We perform single-qudit operations $R_{0k}(\phi,\theta)$ on a particular ion by applying laser pulses resonant to the transitions between corresponding qudit states.
The corresponding operation matrices have the following form:
\begin{equation}
	R_{01}(\phi, \theta){=}\left(
	\begin{array}{cccc}
		\cos(\frac{\theta}{2})   &   -ie^{-i\phi}\sin(\frac{\theta}{2})      &       0       &       0\\
		-ie^{i\phi}\sin(\frac{\theta}{2}) & \cos(\frac{\theta}{2}) & 0 & 0\\
		0 & 0 & 1 & 0\\
		0 & 0 & 0 & 1
	\end{array} \right)\,,
\end{equation}
\begin{equation}
	R_{02}(\phi, \theta){=}\left(
	\begin{array}{cccc}
		\cos(\frac{\theta}{2})   &   0      &       -ie^{-i\phi}\sin(\frac{\theta}{2})       &       0\\
		0 & 1 & 0 & 0\\
		-ie^{i\phi}\sin(\frac{\theta}{2}) & 0 & \cos(\frac{\theta}{2}) & 0\\
		0 & 0 & 0 & 1
	\end{array}
\right)\,,
\end{equation}
	\begin{equation}
	R_{03}(\phi, \theta){=}\left(
	\begin{array}{cccc}
		\cos(\frac{\theta}{2})   &   0      &      0       &        -ie^{-i\phi}\sin(\frac{\theta}{2})\\
		0 & 1 & 0 & 0\\
		0 & 0 & 1 & 0\\
		-ie^{i\phi}\sin(\frac{\theta}{2}) & 0 & 0 & \cos(\frac{\theta}{2})
	\end{array}
\right)\,,
\end{equation}
where $\theta=\Omega\tau_S$ ($\Omega$ is the Rabi-frequency and $\tau_S$ is the laser pulse duration) and $\phi$ is the phase, which can be controlled by the laser field phase.

To entangle two particles we use the well-known M$\o$lmer-S$\o$rensen gate~\cite{Blatt2003-2,Molmer-Sorensen1999,Molmer-Sorensen1999-2,Molmer-Sorensen2000} with $\pi/4$ phase.
We apply this operation to $|0\rangle$ and $|1\rangle$ states of the neighbouring ions, so further we denote it as  $XX_{01,01}^{AB}(\pi/4)$.  
If we neglect additional phases gained by each of the qudits~\cite{Ringbauer2021}, 
as they can be compensated by a set of single-qudit operations, the gate can be described by the matrix:
\begin{equation} 
\begin{aligned}
&XX_{01,01}^{AB}(\chi)=\exp\left(-i\chi \tilde{X}_{01} \otimes \tilde{X}_{01}\right), \\ 
&\tilde{X}_{01}=\left(
\begin{array}{cccc}
0 & 1 & 0 & 0\\
1 & 0 & 0 & 0\\
0 & 0 & 0 & 0\\
0 & 0 & 0 & 0\\
\end{array}
\right)\,,
\end{aligned}
\end{equation}
where $\chi$ is the phase. We note that $XX_{01,01}^{AB}(\pi/4)$ acting on $|00\rangle^{AB}$ leads to the generation of the Bell state having the following form:
\begin{equation}\label{eq:bell}
    |\Psi\rangle^{AB}=\left(|00\rangle^{AB}-i|11\rangle^{AB}\right)/\sqrt{2}.
\end{equation}

To perform the MS-gate, we apply a bi-chromatic laser field with frequencies of $\omega_{01}\pm(\omega_m+\delta)$ onto both ions, 
where $\omega_{01}$ is the resonant frequency of the $|0\rangle\to|1\rangle$ transition, $\omega_m$ is the vibrational mode frequency used for entanglement, and $\delta$ is the detuning. 
To achieve the required gate phase, the pulse parameters must satisfy equation $\tau_{\rm MS}= 2\pi/\delta=\pi/\eta\Omega$, 
where $\eta$ is the Lamb-Dicke parameter for the selected mode and $\Omega$ is the resonant Rabi-frequency of $|0\rangle\to|1\rangle$ transition. 
To entangle ions, we use the axial stretch mode with $\omega_m=2\pi \times 809$ kHz. 
Axial modes are better spectrally separated from each other with respect to the radial ones; this enables one to neglect interactions with other modes. 
The stretch mode was chosen over the center-of-mass mode due to its higher frequency and lower sensitivity to heating, which reduces, first, the influence of the laser phase noise and, second, affect of ions anomalous heating on the gate fidelity.
It can be shown (see e.g., Ref.~\cite{Ringbauer2021}) that this gate set, which consists of single-qudit operations and a two-qudit MS gate, is universal for this system.

\subsection{Qudit readout}

The readout process for qudits is more complicated than for qubits, because more than two states should be distinguished. 
For a full readout of $^{171}$Yb$^{+}$ ion qudits one may use the following approach.
First, similarly to the usual electron shelving method~\cite{bergquist1986observation}, in order to project the ion onto $|0\rangle$ or a subspace spanned by other states we apply cooling and repumping laser (without modulation at 3.07 GHz) beams. 
This results in either strong fluorescence or its absence depending on the projection result, which can be distinguished with a photomultiplier tube (PMT) or an EMCCD camera.
The population from $|0\rangle$ in this process is pumped to the $\,^2S_{1/2}(F=1)$ state. 
Then a $R_{0k}(0,\pi)$ gate is applied for transferring the population from $|k\rangle$ into $|0\rangle$, and the measurement is repeated. 
By performing the above procedure successively for each of the upper qudit states, the full readout is completed. 
For simplicity, one can omit reading out one of the states, as its population can be calculated from all the other measurements. 

In order to reduce the degradation of the readout fidelity due to multiple imperfect single-qudit operations, 
we measure the population of only one state in each experimental run, and then repeat the experiment again to measure others.  
Using this approach, the probability to correctly distinguish $|0\rangle$ state from others is the same as for optical qubits~\cite{semenin2021optimization}. 
For other states it is reduced due to one additional single-qudit operation.
We note that the used sequential readout method is not scalable, but it is useful for our proof-of-principle demonstration.
Our plan is to swith to the first described readout method in the new version of our experimental setup.

\section{Trapped-ion quantum processor with optical qudits}\label{sec:setup}

\begin{figure*}[ht!]
	\includegraphics[width=1\linewidth]{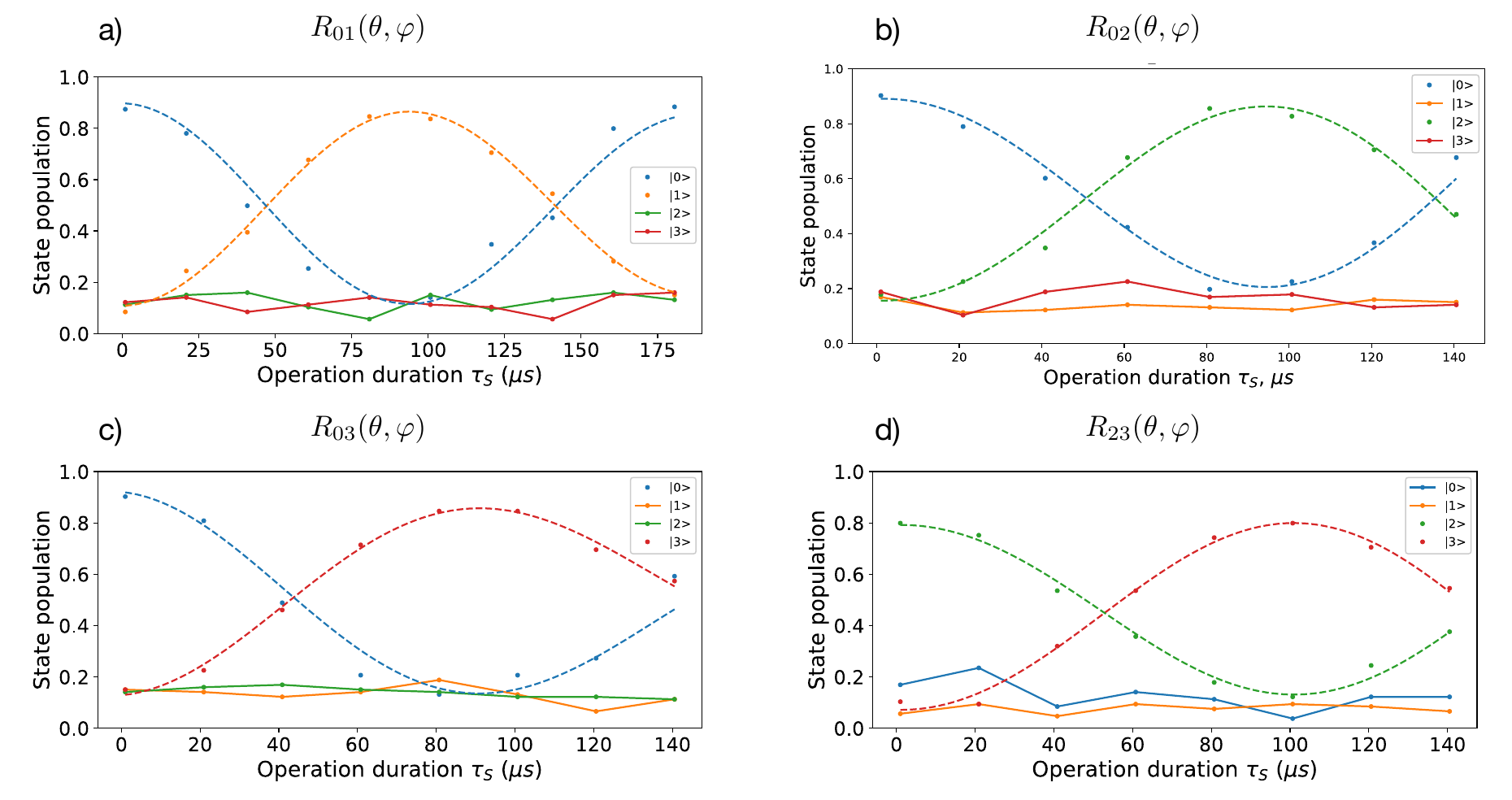}
	\vskip-3mm
	\caption{Single-qudit gate realizations:
	dynamics of state populations when performing elementary single-qudit gates (a) $R_{01}(\theta,\varphi)$, (b) $R_{02}(\theta,\varphi)$, (c) $R_{03}(\theta,\varphi)$, and a composite gate (d) $R_{23}(\theta,\varphi)$ when gate duration $\tau_S \propto \theta$ is scanned. In the latter case ion was initially prepared into the $|2\rangle$ state by the $\pi$-pulse on the $|0\rangle\rightarrow|2\rangle$ transition, after which a composite gate $R_{23}(\theta,\varphi)=R_{02}(\pi,\pi/2)R_{03}(\theta,\varphi)R_{02}(\pi,-\pi/2)$ followed. }
	\label{fig:singe-qudit}
\end{figure*}

Here we describe our experimental setup for realizing a trapped-ion quantum processor with optical qudits.
For trapping $^{171}$Yb$^{+}$ ions we use an one-segment linear 4-blades Paul trap with the secular frequencies of $\{\omega_x,\omega_y,\omega_z\}=2\pi \times \{1520,1500,467\}$ kHz (see Fig.~\ref{fig:trap}).
The upper levels' degeneracy is lifted up by applying a homogenous magnetic field of $B=500~\mu$T, which is also necessary for efficient laser cooling of $^{171}$Yb$^{+}$ ions by destabilization of the coherent ``dark states''~\cite{Arimondo1996}. 
The Zeeman splitting of the transition in this case appears to be 4.2 MHz. 
The magnetic field is produced by three pairs of orthogonally aligned coils with no special magnetic shielding.

We realize a two-particle ion string and use four states in each ion, $^2S_{1/2}(m_F=0)$ and $^2D_{3/2}(m_F=0,\pm1)$, for encoding two ququarts, which we further refer to as qudit A and qudit B. 
At the beginning of each experimental run ions are Doppler cooled for 5 ms, which brings the temperature of the ion crystal down to 1.7 mK~\cite{ZalivakoSemerikov2021}. 
Doppler cooling is achieved with diode lasers at 369.5 nm and 935.2 nm with electro-optical modulators (EOMs) at 14.7 GHz and 3.07 GHz, respectively, to avoid population trapping in metastable hyperfine components. 
It is followed by qudit initialization to $|0\rangle$ by switching off the 14.7 GHz EOM and switching on the 2.1 GHz EOM installed in the 369.5 nm beam. 
The pumping process takes 5 $\mu$s.

Qudit manipulation on the quadrupole $\,^2S_{1/2}(F=0,m_F=0) \to \,^2D_{3/2}(F=2,m_F=k)$ transitions is performed with a frequency doubled diode master-oscillator power-amplified laser system generating at fundamental wavelength of 871 nm. 
The oscillator is locked to a high-finesse external optical cavity made from ultra-low-expansion (ULE) glass providing the spectral linewidth of less than 30 Hz~\cite{Kolachevsky2018}.
The linewidth was directly measured by spectroscopy of the qudit transition. 
The laser beam at 435.5 nm is split into two parts, one of which addresses ions globally along the trap axis and is used for two-qudit gates. 
The second part illuminates ions orthogonally to the trap axis and enables one to address ions individually. 
To address a particular ion, a pair of acousto-optic deflector (AOD) and acousto-optic modulator (AOM) is used. 
The former allows spatial scanning of the addressing beam through ions, while the latter is used for the beam frequency tuning and fast switching.
We note that although the global addressing for entangling operations can be used for proof-of-principle demonstration, it prevents scaling of this system to larger number of ions. In following experiments we plan to replace it with two individual addressing beams orthogonal to the trap axis and, thus using radial modes of motion for particles entanglement. Although it requires more complex modulation of the addressing beams due to small spectral separation of radial modes, it is a well established approach for scalable manipulation for systems of dozens of ions within the trap \cite{Monroe2016, Monroe2019, Blatt2021-3, blumel2021power}.

After the state initialization, a resolved-sideband technique~\cite{Monroe1995} via $|0\rangle \to |1\rangle$ transition is used to achieve a ground state cooling of both axial modes of ion motion. 
The mean number of phonons in the stretch mode, used for ions entanglement, achieves $n_{\rm st}=0.079\pm0.013$. 

The ground state cooling is followed by the quantum operations themselves and readout. 
Ions fluorescence for readout is collected with a single aspheric lens which allows for individual ions resolving and is sent to the detector (PMT or EMCCD camera, depending on the experiment). 

\section{Realization of the single-qudit and two-qudit gates}\label{sec:results}

Here we present results of proof-of-principle experiments with optical trapped ion qudits. 
We start with realizing single-qudit gates and estimating their performance.
To estimate the fidelity of the single-qudit $R_{0k}(\phi,\theta)$ operations, we excite Rabi-oscillations between $|0\rangle$ and $|k\rangle$ states for $k=1,2,3$ with the pulse durations up to 200 $\mu$s (see Fig.~\ref{fig:singe-qudit}).
The resulting data is fitted with an exponentially damped sine function. 
The corresponding gate fidelity is estimated as the population of the $|k\rangle$ state in the first maximum, which is averaged over 300 measurements for each pulse duration. 

The fidelities for elementary single-qudit gates obtained with two qudits in the trap are presented in Table~\ref{tab:single-qudit}; they range from $83\%$ to $89\%$. 
At the moment our fidelities are limited by the temperature of the radial ion modes, which are not ground-state cooled. 
Readout errors also contribute to the estimated gate infidelities and, in particular, cause specious spectator qudit states excitations.
The fidelity appears to be lower than in the optical qubit experiments (94~\%) reported in Ref.~\cite{ZalivakoSemerikov2021}. 
This can be explained by excessive heating of radial modes (above the Doppler limit) during the ground state cooling of axial modes. 
After ground state cooling of all vibrational modes in a next-generation setup, we expect the single-qudit fidelities to be at the same level as for $^{40}$Ca$^{+}$.

\begin{table}[]
\begin{tabular}{|l|l|l|l|l|}
\hline
        & 0$\to$1                     & 0$\to$2                     & 0$\to$3          & 2$\to$3   \\ \hline
Qudit A & $\mathcal{F}_{01}^{A}=85\%$ & $\mathcal{F}_{02}^{A}=83\%$ & $\mathcal{F}_{03}^{A}=87\%$ & $\mathcal{F}_{23}^{A}=65\%$ \\ \hline
Qudit B & $\mathcal{F}_{01}^{B}=87\%$ & $\mathcal{F}_{02}^{B}=89\%$ & $\mathcal{F}_{03}^{B}=87\%$ & $\mathcal{F}_{23}^{B}=70\%$ \\ \hline
\end{tabular}
\caption{Fidelities of single-qudit gates for two ion qudits. Operations were performed with two ions in the trap. 
The operation 2$\to$3 is composed from three elementary ones.}
\label{tab:single-qudit}
\end{table}

Non-point focusing of our optical addressing system leads to the excitation of the second ion while addressing the first one, which can be interpreted as a cross-talk. 
We measure the cross-talk by reading out the Rabi-oscillations of one of the ions while exciting the other one by the laser field. 
The cross-talk turns out to be smaller than 10\% and is due to the infidelity of the addressing optics. 
The discussion of the state preparation and measurement (SPAM) errors for our system in may be found in Ref.~\cite{ZalivakoSemerikov2021}. 

We also demonstrated the set composite of $R_{ik}(\phi,\theta)$ operations ($i,j=1,2,3; i \neq j$) by applying consequential $R_{0k}$ and $R_{0i}$ pulses.
The measured fidelity matches with what is expected to obtain from the corresponding product of each elementary operations.  

\begin{figure}
\center{\includegraphics[width=1\linewidth]{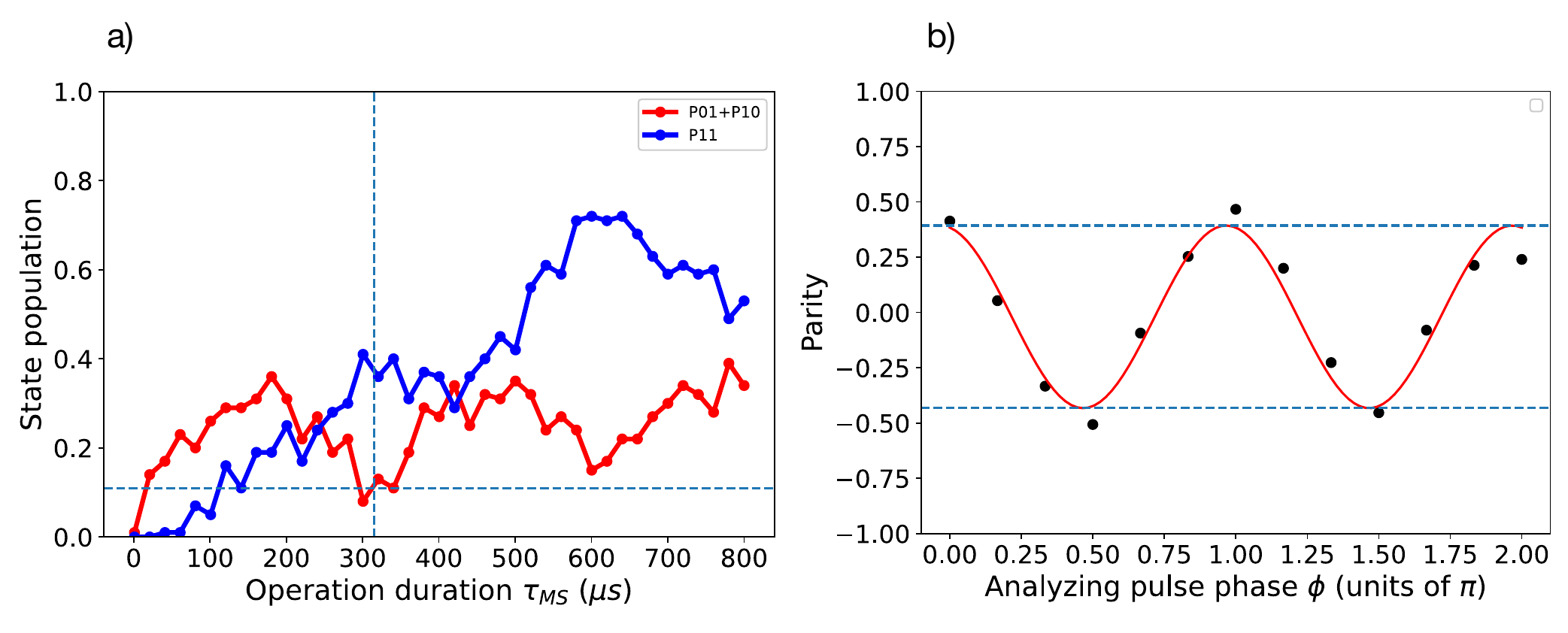}}
	\vskip-3mm
	\caption{Two-qudit gate realization:
	(a) probability of finding 1 ($P_{01}+P_{10}$) or 2 ($P_{11}$) ions in an excited state $|1\rangle$ after applying a bi-chromatic laser pulse of duration $\tau_{MS}$; vertical dotted line shows $\tau_{MS}$ corresponding to $XX_{01,01}^{AB}(\pi/4)$ gate;
	(b) parity (\ref{eq:parity}) oscillations $P_a(\phi)$ after applying a two-qudit gate $XX_{01,01}^{AB}(\pi/4)$ and analyzing gate $R_{01}^{AB}(\phi,\pi/2)$. Initially ions were prepared in $|00\rangle^{AB}$ state.}
	\label{fig:main}
\end{figure}

Fidelity of the two-qudit operation was estimated by following Refs.~\cite{Wineland2003,Blatt2008}, i.e. by measuring fidelity of the Bell state preparation. 
Two experiments were performed. 
In the first, we scan the duration of the bi-chromatic laser pulse for the  entangling operation, 
after ions were initialized into $|00\rangle$. After that probabilities $P_{00}$, $P_{01} + P_{10}$, $P_{11}$ of finding 0, 1, or 2 ions in the state $|1\rangle$ (see Fig.~\ref{fig:main}b) are measured. 
The optimal gate duration time manifests itself as the first dip in $P_{01} + P_{10}$. 
The magnitude of $P_{01} + P_{10}=1-\rho_{00,00}+\rho_{11,11}$ at this point characterizes the diagonal terms of the prepared state density matrix (here and after, $\rho_{ij,kl}={\rm Tr}\left(\rho|i\rangle\langle j|\otimes|k\rangle \langle l|\right)$).

To estimate non-diagonal density matrix elements of the prepared Bell state, parity oscillations are measured. 
We apply $XX_{01,01}^{A,B}(\pi/4)$ gate with the optimal duration, after which the global $R_{01}(\phi,\pi/2)$ analyzing gate is performed. 
The dependency of the parity $P_a=P_{11}+P_{00}-P_{01}-P_{10}=1-2(P_{01}-P_{10})$ on the analyzing pulse phase $\phi$ is measured (see Fig.~\ref{fig:main}c). 
Fitting this dependence with function 
\begin{equation}\label{eq:parity}
\begin{aligned}
    P_a(\phi)=A\sin2(\phi+\phi_0), 
\end{aligned}
\end{equation}
where $A$ and $\phi_0$ are free parameters, allows one to determine the coherence part of the prepared state density matrix as  $|\rho_{00,11}|=A/2$.

The overall Bell-state preparation fidelity can be estimated as follows:
\begin{equation}
\begin{aligned}
	\mathcal{F}_{\rm Bell}&=\langle{\Psi^{AB}|\rho|\Psi^{AB}}\rangle \\
	&=(\rho_{00,00}+\rho_{11,11})/2+|\rho_{00,11}|.
\end{aligned}
\end{equation}

Using such an estimation, we obtain $\mathcal{F}_{\rm MS}=65\pm4\%$ for two-qudit entanglement gate with the duration of $\tau_{\rm MS}=310$ $\mu$s.
The main source of the two-qudit gate infidelity in our setup is the spectral phase noise of the addressing laser, which excite the carrier transition during the gate, and high heating rates of the trap. 
As we observe, the two-qudit gate fidelity strongly depends on the parameters of the lock of the addressing laser to the cavity. 
As the lock parameters determine the noise spectrum of the laser, we can assume that it has a leading contribution to the gate error.

\begin{figure}
\center{\includegraphics[width=1\linewidth]{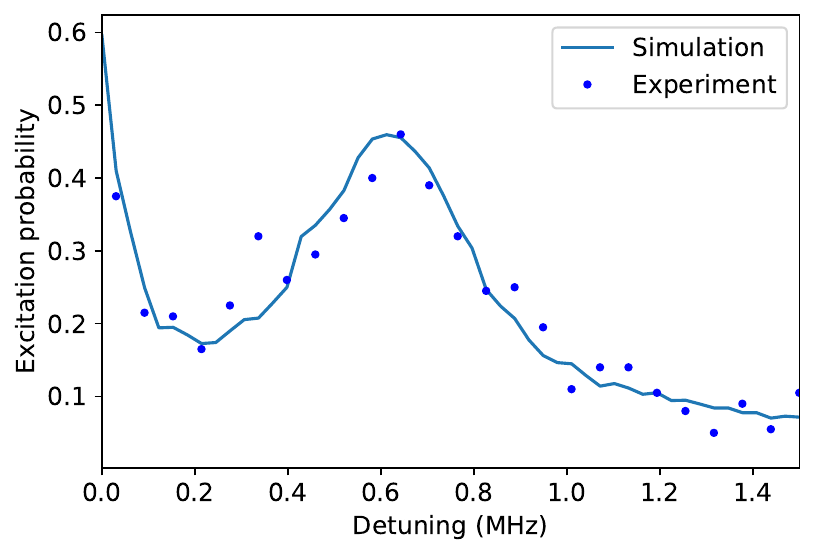}}
	\vskip-3mm
	\caption{Dependence of the $|0\rangle\to|1\rangle$ transition excitation probability on the laser detuning from the resonance (experimental data and simulation results). 
	Rabi frequency is $\Omega = 2\pi\times 45$ kHz, laser pulse duration $\tau_{p}=400 \,\mu s$. 
	As at detunings larger than $\Omega/2\pi$ excitation is mostly determined by the laser phase noise, fitting experimental data with a numeric model enables one to extract laser noise parameters.}
	\label{fig:noise_fit}
\end{figure}

To further validate this hypothesis we performed a numeric simulation of the experiment including laser phase noise influence following \cite{nakav2022effect}. 
To retrieve a laser noise power spectral density (PSD) an ion spectroscopy was carried out on a $|0\rangle\to|1\rangle$ transition with an interrogating pulse duration much longer than a $\pi$-pulse. 
Ion excitation at detunings larger than an on-resonance Rabi frequency is then determined by noise PSD at this detuning. 
Varying PSD in the numeric simulation and fitting experimental data with a simulated curve allows one to extract information about the laser phase noise. 
In Fig. ~\ref{fig:noise_fit} both measured spectrum and a simulated curve for optimized PSD parameters are shown.  After that extracted PSD data was used to simulate MS gate. 
The fidelity appeared to be on the level of $70\%$ which is close to experimentally observed results and proves laser noise to be the major contribution to the gate error.
Thus, as we expect, existing shortcomings that prevent us from the demonstration of the state-of-the-art fidelities of quantum gates with ytterbium ions are of technical nature. 

\section{Conclusion and outlook}\label{sec:conclusions}

We have presented the proof-of-principle realization of the two-ququarts quantum processor based on  $^{171}$Yb$^{+}$ ions. 
Such a system is computationally analogous to a four-qubit processor and, thus, it may indicate on the potential advantages of the qudit approach to quantum computing. 
The qudits are encoded in the Zeeman structure of $\,^2S_{1/2}(F=0)$ and $\,^2D_{3/2}(F=2)$ levels, which are coupled via electric quadrupole transition at 435.5 nm. 
The system allows encoding qudits with $d$ up to 6, however in this demonstration only part of the available levels were used ($d=4$). 
We have realized single-qudit and two-qudit gates which constitute a universal gate set. The estimated fidelities of single-qudit gates range from $83\%$ to $89\%$ and $65\%$ for two-qudit gate. 
These fidelities are limited at the moment by the temperature of the ions and addressing laser phase noise. 
The results can be improved in our next generations of the setup, so we expect to reach the parameters regime reported in e.g. in Ref.~\cite{Ringbauer2021}. 
At the same time, the energy structure of the $^{171}$Yb$^{+}$ promises several advantages in developing decoherence-free qudit systems~\cite{Valahu2022}.

As we expect, with the increase in the number of qudits and fidelity of operations, qudit-based quantum processors will be able to demonstrate their advantages in realizing quantum algorithms in comparisons with their qubit-based counterparts. 
As it has been shown in Ref.~\cite{Nikolaeva2021}, the qudit-based realization provides an advantage both in the circuit width and depth.
We would also like to note that the use of qudits besides quantum computing, offers certain perspectives in quantum teleportation~\cite{Pan2019} and quantum communications~\cite{Gisin2002,Boyd2015}, 
as well as opens up opportunities for uncovering fundamental concepts of quantum mechanics~\cite{Li2013,Frydryszak2017,Zyczkowski2022}.
Thus, we expect that the future research on trapped ion qudits will be beneficial for various research avenues. 

\section*{Acknowledgements}

We thank A.S. Nikolaeva, E.O. Kiktenko, and L.R. Bakker for fruitful discussions and useful comments.
The experimental work was supported by Leading Research Center on Quantum Computing (Agreement No. 014/20) and Russian Roadmap on Quantum Computing (Contract No. 868- 1.3-15/15-2021, October 5, 2021).
The theoretical work of A.K.F. (analysis in Sec.~\ref{sec:optqudit} and~\ref{sec:results}) is also supported by the RSF grant 19-71-10092.

\bibliography{bibliography.bib}

\end{document}